\documentclass[12pt]{article}

\usepackage{graphicx,cite}
\usepackage{epsf,latexsym}
\usepackage{amssymb,amsmath}
\usepackage{dcolumn}
\renewcommand{\baselinestretch}{1.5}

\begin{document}

\def\llra{\relbar\joinrel\longrightarrow}              
\def\mapright#1{\smash{\mathop{\llra}\limits_{#1}}}    
\def\mapup#1{\smash{\mathop{\llra}\limits^{#1}}}     
\def\mapupdown#1#2{\smash{\mathop{\llra}\limits^{#1}_{#2}}} 

\title{\bf Spontaneous fission half-lives of heavy and superheavy nuclei within a generalized liquid drop model}

\author{Xiaojun Bao$^{1}$, Hongfei Zhang$^{1}$\thanks{Corresponding author. Tel.: +86 931
8622306. E-mail address: zhanghongfei@lzu.edu.cn}, G. Royer$^{2}$,
Junqing Li$^{1,3}$}

\maketitle
\begin{center}
\begin{enumerate}
\item School of Nuclear Science and Technology, Lanzhou University, Lanzhou 730000, China
\item Laboratoire Subatech, UMR: IN2P3/CNRS-Universit\'{e}-Ecole des Mines, 4 rue A. Kastler, 44 Nantes, France
\item Institute of Modern Physics, Chinese Academy of Science, Lanzhou 730000, China
\end{enumerate}
\end{center}

\maketitle

\begin{abstract}
\noindent We systematically calculate the spontaneous fission
half-lives for heavy and superheavy nuclei between U and Fl
isotopes. The spontaneous fission process is studied within the
semi-empirical WKB approximation. The potential barrier is
obtained using a generalized liquid drop model, taking into
account the nuclear proximity, the mass asymmetry, the
phenomenological pairing correction, and the microscopic shell
correction. Macroscopic inertial-mass function has been employed
for the calculation of the fission half-life. The results
reproduce rather well the experimental data. Relatively long
half-lives are predicted for many unknown nuclei, sufficient to
detect them if synthesized in a laboratory.
\end{abstract}


\newpage
\maketitle
\section{Introduction}\label{intro}\noindent
Spontaneous fission of heavy nuclei was first predicted by Bohr
and Wheeler in 1939 \cite{N1J39}. Their fission theory was based
on the liquid drop model. Interestingly, their work also contained
an estimate of a lifetime for fission from the ground state. Soon
afterwards, Flerov and Petrzak \cite{F2P40} presented the first
experimental evidence for spontaneous fission. Since this
discovery of spontaneous fission of $^{238}$U, fission of numerous
other actinide nuclei have been reported experimentally
\cite{N3D00}. Several theoretical approaches, both
phenomenologically and microscopically, can be employed to
investigate spontaneous fission. Early descriptions of fission
were based on a purely geometrical framework of the charged liquid
drop model \cite{N1J39}. In 1955 Swiatecki \cite{W4J55} suggested
that more realistic fission barriers could be obtained by adding a
correction energy to the minimum in the liquid drop model barrier.
The correction was calculated as the difference between the
experimentally observed nuclear ground-state mass and the mass
given by the liquid drop model. Swiatecki obtained much improved
theoretical spontaneous fission half-lives based on these modified
liquid drop model barriers. These observations formed the basis
for the shell-correction method. In the mid-1960s, Strutinsky
\cite{V5M67,V6M68} presented a method to theoretically calculate
these shell corrections. Quantum shell effects are added to the
average behavior described by the liquid drop. This
macroscopic-microscopic approach turned out to be very successful
in explaining many features of spontaneous fission
\cite{P7M01,P8M09,A9S07}. As compared to $\alpha$-decay, the
spontaneous fission is much more complex and there are some data
such as mass and charge numbers of the two fragments, number of
emitted neutrons and released energy, ...which are very difficult
to reproduce. The full microscopic treatment of such a
multidimensional system is extremely complex. In particular, a
microscopic calculation of spontaneous fission half-lives is very
difficult due to both the complexity of the fission process and
the uncertainty on the height and shape of the fission barrier
\cite{P10J87,P11J89}. Indeed, the deformation energy of the
nucleus undergoes a significant change when the nuclear shape
turns into a strongly deformed configuration of two fragments in
contact at the scission point. Furthermore, it is known that the
spontaneous fission half-life is very sensitive to small changes
of the various quantities appearing in the calculations.

Spontaneous fission is one of the most prominent decay modes,
energetically feasible for heavy and superheavy nuclei(SHN).
Recently, the spontaneous fission half-lives of several SHN have
been measured by different laboratories
\cite{K12E06,J13D06,D14P06,Y15T05,Y16T06,Y17T04}. The fission as
well as $\alpha$-decay probability determines the stability of
these newly synthesized SHN.

Within a Generalized Liquid Drop Model (GLDM) taking into account
the mass and charge asymmetry and the proximity energy, the
deformation energy of compact and creviced shapes have been
determined. The proximity forces strongly lower the
 deformation energy of these quasimolecular shapes and the calculated fission barrier heights agree well with
the experimental results \cite{G18B84,G19F95,G20K02,C21G06}.
Within the same approach, the $\alpha$-decay
\cite{G22R00,H23Z06,G24H08,H25F08}, cluster emission \cite{G26R01}
and fusion \cite{G27B85} data can also be reproduced.

The present work is closely connected with the intensive
experimental activity on the synthesis and study of heavy and SHN
in recent years. It aims at the interpretation of existing
experimental fission half-lives and in predictions of half-lives
of yet unknown nuclei. By using the GLDM and taking into account
the ellipsoidal deformations of the two different fission
fragments and their associated microscopic shell corrections and
pairing effects, we have systematically calculated the spontaneous
fission half-lives of nuclei in the mass region from $^{232}$U to
$^{286}$Fl by considering all the possible mass and charge
asymmetries.

The study is focused on quasimolecular shapes since these shapes
have been rarely investigated in the past. Indeed, this
quasimolecular shape valley where the nuclear proximity effects
are so important is inaccessible in using the usual development of
the nuclear radius. Furthermore, microscopic studies within HFB
theory cannot describe the evolution of two separated fragments
and cannot simulate the formation of a deep neck between two
almost spherical fragments. The reproduction of the proximity
energy is also not obvious within mean-field approaches and the
definition of the scission point is not easy for very elongated
shapes.

The paper is organized as follows. The selected shape sequence and
macroscopic-microscopic model are described in Section 2, results
and discussions are given in Section 3, and conclusions and
summary are presented in Section 4.
\section{Shape sequence and macroscopic-microscopic model}\label{model}\noindent
\subsection{Quasimolecular shapes}
The shape is given simply in polar coordinates (in the plane
$\phi=0$) by \cite{G28B82}
\begin{equation}R(\theta)^{2}=
\begin{cases}a^{2}\sin^{2}\theta+c_{1}\cos^{2}\theta
  ~~(0\leq\theta\leq \pi/2) \\a^{2}\sin^{2}\theta+c_{2}\cos^{2}\theta
  ~~(\pi/2\leq\theta\leq \pi)
\end{cases}
\end{equation}
where $c_{1}$ and $c_{2}$ are the two radial elongations and $a$
is the neck radius. Assuming volume conservation, the two
parameters $s_{1}=a/c_{1}$ and $s_{2}=a/c_{2}$ completely define
the shape. The radii of the future fragments allow to connect
$s_{1}$ and $s_{2}$:
\begin{eqnarray}
s_{2}^{2}=\frac{s_{1}^{2}}{s_{1}^{2}+(1-s_{1}^{2})(R_{2}/R_{1})^{2}}.
\end{eqnarray}
When $s_{1}$ decreases from 1 to 0 the shape evolves continuously
from one sphere to two touching spheres with the natural formation
of a deep neck while keeping almost spherical ends. So, we would
like to point out that the most attractive feature of the
quasimolecular shapes is that it can describe the process of the
shape evolution from one body to two separated fragments in a
unified way.

\subsection{GLDM energy}
Within GLDM the macroscopic energy of a deformed nucleus is
defined as
\begin{eqnarray}
E=E_{V}+E_{S}+E_{C}+E_{prox},
\end{eqnarray}
where the different terms are respectively the volume, surface, Coulomb and nuclear proximity energies. \\
For one-body shapes, the volume $E_{V}$, surface $E_{S}$ and
Coulomb $E_{C}$ energies are given by
\begin{eqnarray}
E_{V}=-15.494(1-1.8I^{2})A  ~MeV,
\end{eqnarray}
\begin{eqnarray}
E_{S}=17.9439(1-2.6I^{2})A^{2/3}(S/4\pi R_{0}^{2}) ~MeV,
\end{eqnarray}
\begin{eqnarray}
E_{C}=0.6e^{2}(Z^{2}/R_{0})B_{C}.
\end{eqnarray}
$B_{C}$ is the Coulomb shape dependent function, $S$ is the
surface and $I$ is the relative neutron excess.
\begin{eqnarray}
B_{C}=0.5\int(V(\theta)/V_{0})(R(\theta)/R_{0})^{3}\sin\theta
d\theta,
\end{eqnarray}
where $V(\theta)$ is the electrostatic potential at the surface
and $V_{0}$ is the surface potential of the sphere. The effective
sharp radius $R_{0}$ has been chosen as
\begin{eqnarray}
R_{0}=(1.28A^{1/3}-0.76+0.8A^{-1/3}) ~fm .
\end{eqnarray}
This formula proposed in Ref. \cite{J29J77} is derived from the
droplet model and the proximity energy and simulates rather a
central radius for which $R_{0}/A^{1/3}$ increases slightly with
the mass. It has been shown \cite{G22R00,G27B85} that this
selected more elaborated expression can also be used
to reproduce accurately the fusion, fission, cluster and alpha decay data.\\

For two-body shapes, the coaxial ellipsoidal deformations have
been considered \cite{G30C92}. The system configuration depends on
two parameters: the ratios s$_{i}$ ($i=1,2$) between the
transverse semi-axis $a_{i}$ and the radial semi-axis $c_{i}$ of
the two different fragments
\begin{eqnarray}
a_{i}=R_{i}s_{i}^{1/3}~~and~~c_{i}=R_{i}s_{i}^{-2/3}.
\end{eqnarray}
The prolate deformation is characterized by s$\leq$1 and the
related eccentricity is written as $e^{2}=1-s^{2}$ while in the
oblate case s$\geq$1 and $e^{2}=1-s^{-2}$. The volume and surface
energies are $E_{V_{12}}=E_{V_{1}}+E_{V_{2}}$ and
$E_{S_{12}}=E_{S_{1}}+E_{S_{2}}$. In the prolate case, the
relative surface energy reads
\begin{eqnarray}
B_{Si}=\frac{(1-e_{i}^{2})^{1/3}}{2}[1+\frac{\sin^{-1}(e_{i})}{e_{i}(1-e_{i}^{2})^{1/2}}]
\end{eqnarray}
and in the oblate case
\begin{eqnarray}
B_{Si}=\frac{(1+\epsilon_{i}^{2})^{1/3}}{2}[1+\frac{\ln(\epsilon_{i}+(1+\epsilon_{i}^{2})^{1/2})}
{\epsilon_{i}(1+\epsilon_{i}^{2})^{1/2}}],
\end{eqnarray}
where $\epsilon_{i}^{2}=s_{i}^{2}-1$.

The Coulomb self-energy of the spheroid $i$ is
\begin{eqnarray}
E_{C,self}=\frac{3e^{2}Z_{i}^{2}B_{ci}}{5R_{i}}.
\end{eqnarray}
The relative self-energy is, in the prolate case
\begin{eqnarray}
B_{Ci}=\frac{(1-e_{i}^{2})^{1/3}}{2e_{i}}\ln\frac{1+e_{i}}{1-e_{i}}
\end{eqnarray}
and in the oblate case
\begin{eqnarray}
B_{Ci}=\frac{(1+\epsilon_{i}^{2})^{1/3}}{\epsilon_{i}}\tan^{-1}\epsilon_{i}.
\end{eqnarray}
The Coulomb interaction energy between the two fragments reads
\begin{eqnarray}
E_{C,int}=\frac{e^{2}Z_{1}Z_{2}}{r}[s(\lambda_{1})+s(\lambda_{2})-1+S(\lambda_{1},\lambda_{2})]
\end{eqnarray}
where $\lambda_{i}^{2}=(c_{i}^{2}-a_{i}^{2})/r^{2}$, $r$ is the
distance between the two mass centers.

In the prolate case, s$(\lambda_{i})$ is expressed as
\begin{eqnarray}
s(\lambda_{i})=\frac{3}{4}(\frac{1}{\lambda_{i}}-\frac{1}{\lambda_{i}^{3}})
\ln(\frac{1+\lambda_{i}}{1-\lambda_{i}})+\frac{3}{2\lambda_{i}^{2}},
\end{eqnarray}
while for the oblate shapes
\begin{eqnarray}
s(\lambda_{i})=\frac{3}{2}(\frac{1}{\omega_{i}}+\frac{1}{\omega_{i}^{3}})
\tan^{-1}\omega_{i}-\frac{3}{2\omega_{i}^{2}},
\end{eqnarray}
where $\omega_{i}^{2}=-\lambda_{i}^{2}$.

$S(\lambda_{1},\lambda_{2})$ can be represented in the form of a
two-fold summation
\begin{eqnarray}
S(\lambda_{1},\lambda_{2})=\sum_{j=1}^{\infty}\sum_{k=1}^{\infty}\frac{3}{(2j+1)(2j+3)}
\\ \nonumber
\frac{3}{(2k+1)(2k+3)}\frac{(2j+2k)!}{(2j)!(2k)!}\lambda_{1}^{2j}\lambda_{2}^{2k}.
\end{eqnarray}

The surface energy results from the effects of the surface tension
forces in a half space. When a neck or a gap appears between
separated fragments an additional term called proximity energy
must be added to take into account the effects of the nuclear
forces between the close surfaces. It moves the barrier top to an
external position and strongly decreases the pure Coulomb barrier:
\begin{eqnarray}
E_{prox}(r)=2\gamma\int_{h_{min}}^{h_{max}}\Phi[D(r,h)/b]2\pi hdh
\end{eqnarray}
where
\begin{eqnarray}
\gamma=0.9517\sqrt{(1-2.6I_{1}^{2})(1-2.6I_{2}^{2})} MeV fm^{-2}.
\end{eqnarray}
$h$ is the transverse distance varying from the neck radius or
zero to the height of the neck border, $D$ is the distance between
the opposite surfaces in consideration and $b$ is the surface
width fixed at 0.99 fm. $\Phi$ is the proximity function. The
surface parameter $\gamma$ is the geometric mean between the
surface parameters of the two fragments.

\subsection{Shell energy}
The shape-dependent shell corrections have been determined within
the Droplet Model expressions \cite{W31D77}:
\begin{eqnarray}
E_{shell}=E_{shell}^{sphere}(1-2.0\alpha^{2})e^{-\alpha^{2}}
\end{eqnarray}
where $\alpha^{2}=(\delta R)^{2}/a^{2}$. The distortion $\alpha$a
is the root mean square of the deviation of the surface from a
sphere, a quantity which incorporates all types of deformation
indiscriminately. The range a has been chosen to be 0.32r$_{0}$.
The whole shell correction energy decreases to zero with
increasing distortion of the nucleus due to the attenuating factor
($e^{-\alpha^{2}}$). The Strutinsky method at large deformations
supposes that the nucleon shells are not affected by the proximity
effects, which is highly improbable when there are a deep neck in
the deformed shape and close surfaces in regard.

The $E_{shell}^{sphere}$ is the shell corrections for a spherical
nucleus,
\begin{eqnarray}
E_{shell}^{sphere}=cE_{sh}
\end{eqnarray}
and is obtained by the Strutinsky method by setting the smoothing
parameter $\gamma=1.15~~\hbar\omega_{0}$ and the order $p=6$ of
the Gauss-Hermite polynomials, where $\hbar\omega_{0}=41A^{-1/3}$
MeV is the mean distance between the gross shells, the sum of the
shell energies of protons and neutrons. Meanwhile, we introduce a
scale factor c to the shell correction. In this work, we choose
$c=0.82$. To obtain the shell correction $E_{shell}^{sphere}$, we
calculate the single-particle levels based on an axially deformed
Woods-Saxon potential and then apply the Strutinsky method. The
single-particle Hamiltonian is written as \cite{W32M10},
\begin{eqnarray}
H=T+V+V_{S.O},
\end{eqnarray}
with the spin-orbit potential
\begin{eqnarray}
V_{S.O}=-\lambda (\frac{\hbar}{2Mc})^{2}\nabla V\cdot
(\vec{\sigma} \times \vec{p}),
\end{eqnarray}
where $M$ is the free nucleonic mass, $\vec{\sigma}$ is the Pauli
spin matrix and $\vec{p}$ is the nucleon momentum.  $\lambda$
means the strength of the spin-orbit potential. Here we set
$\lambda=\lambda_{0}(1+N_{i}/A)$ with $N_{i}=Z$ for protons and
$N_{i}=N$ for neutrons and $\lambda_{0}$ value of 26.3163. The
central potential $V$ is described by an axially deformed
Woods-Saxon form
\begin{eqnarray}
V(\vec{r})=\frac{V_{q}}{1+\exp[\frac{r-R(\theta)}{a}]},
\end{eqnarray}
where the depth $V_{q}$ of the central potential ($q=p$ for
protons and $q=n$ for neutrons) is written as
\begin{eqnarray}
V_{q}=V_{0}\mp V_{S}I,
\end{eqnarray}
with the minus sign for protons and the plus sign for neutrons.
$V_{0}$ and a take the values -47.4784 and 0.7842, respectively.
$V_{S}$ and $I$ are the isospin-asymmetric part of the potential
depth and the relative neutron excess, where
\begin{eqnarray}
V_{S}=c_{sym}[1-\frac{\kappa}{A^{1/3}}+\frac{2-|I|}{2+|I|A}].
\end{eqnarray}
The values of $c_{sym}$ and $\kappa$ are 29.2876 and 1.4492,
respectively \cite{N33M10}.

\subsection{Pairing energy}
The shape-dependent pairing energy has been calculated with the
following expressions of the finite-range droplet model
\cite{P34J93}.

For odd Z, odd N numbers :
\begin{eqnarray}
E_{Pairing}=\frac{4.8B_{S}}{N^{1/3}}+\frac{4.8B_{S}}{Z^{1/3}}-
\frac{6.6}{B_{S}A^{2/3}}.
\end{eqnarray}
For odd Z, even N numbers :
\begin{eqnarray}
E_{Pairing}=\frac{4.8B_{S}}{Z^{1/3}}.
\end{eqnarray}
For even Z, odd N numbers :
\begin{eqnarray}
E_{Pairing}=\frac{4.8B_{S}}{N^{1/3}}.
\end{eqnarray}
For even Z, even N numbers :
\begin{eqnarray}
E_{Pairing}=0.
\end{eqnarray}
The relative surface energy $B_{s}$, which is the ratio of the
surface area of the nucleus at the actual shape to the surface
area of the nucleus at the spherical shape, is given by
\begin{eqnarray}
B_{s}=\frac{\int_{S}dS}{S_{sphere}}.
\end{eqnarray}
The pairing energies  vary with $B_{s}$.

Within this asymmetric fission model the decay constant is simply
given by $\lambda_{i}=\nu_{0}$P$_{i}$, and the assault frequency
$\nu_{0}$ has been taken as $\nu_{0}=10^{20}$s$^{-1}$. The barrier
penetrability P$_{i}$ is calculated within the action integral
\begin{eqnarray}
P_{i}=\exp[-\frac{2}{\hbar}\int_{r_{1}}^{r_{2}}\sqrt{2B(r)(E(r)-E(sphere))}dr].
\end{eqnarray}
The limits of integration $r_{1}$ and $r_{2}$ are the points of
entrance and exit, respectively, into and from the barrier. The
function $B(r)$ is the inertia with respect to $r$ associated with
motion in the fission direction. The fission half-lives have been
calculated within the following semi-empirical model for the
inertia \cite{J341S34}
\begin{eqnarray}
B(r)=\mu(1+k\exp[-\frac{128}{51}(r-R_{sph}/R_{0})])
\end{eqnarray}
where $\mu$ is the reduced mass of the final fragments and $k$ is
a semi-empirical constant, $k=14.8$. $R_{sph}$ is the distance
between the mass centers of the future fragments in the initial
sphere, $R_{sph}/R_{0}=0.75$ in the symmetric case. To obtain the
total fission constant $\lambda$, we have calculated all the
possible spontaneous fission half-lives corresponding to fission
constant $\lambda_{i}$ of the different possible exiting channels
depending on the different mass and charge asymmetries. The
calculated half-lives of all possible $^{234}$U spontaneous
fission channels are shown in Fig.1 versus one of daughter nucleus
mass number for illustration of the method. For given Z$_{1}$ and
Z$_{2}$ values the spontaneous fission half-lives decrease with
one fragment mass number A$_{2}$, reach a minimum and then
increase with increasing A$_{2}$. The total fission constant is
$\lambda$=$\lambda_{1}+\lambda_{2}....+\lambda_{n}$ and the
half-life is finally obtained by $T_{1/2}=(\ln2)/\lambda$.

\begin{figure}[htbp]
\includegraphics[width=0.90\textwidth]{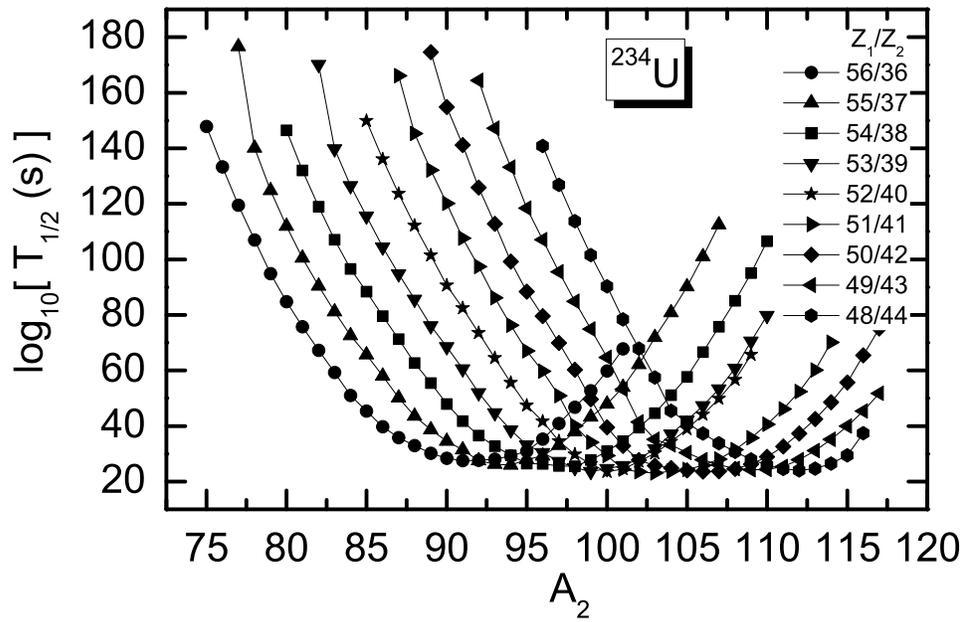}
\caption{\label{Fig1} Calculated spontaneous fission half-lives
for $^{234}$U as a function of one of daughter nucleus mass number.
Parabolic half-live curves represent the different Z$_{1}$ and
Z$_{2}$ combinations, where Z$_{1}$ and Z$_{2}$ indicate the
proton numbers of the two fragments for $^{234}$U spontaneous
fission. }
\end{figure}

\section{Results and discussions}\noindent

The spontaneous fission half-lives of nuclei from $^{232}$U to
$^{286}$Fl have been systematically calculated by using the GLDM
taking into account the microscopic shell corrections and the
shape-dependent pairing energy. The results are listed in Table
\ref{tab:fis}. The first and fourth columns indicate the
spontaneous fissioning nuclei. The experimental
\cite{Y17T04,J44K05} and theoretical spontaneous fission
half-lives are compared in the other columns. The spontaneous
fission half-lives vary in an extremely wide range from $10^{26}$
seconds to $10^{-3}$ seconds when the nucleon number varies from
A=232 to A=286. For a variation nucleon number less than 60, the
amplitude of variation of spontaneous fission half-lives is as
high as $10^{29}$. This leads to the extreme sensitivity of the
half-lives to nucleon number. Thus it is a very difficult task to
reproduce the experimental data accurately. However, the
theoretical spontaneous fission half-lives are in good agreement
with the experimental ones. Among 47 nuclei, 37 experimental
half-lives can be reproduced within a factor of $10^{2}$. Only for
7 nuclei, the deviations between the experimental and theoretical
half-lives are larger than a factor of $10^{3}$. Here the
logarithm of average deviations for 47 spontaneous fission nuclei
is
S=$\sum_{i=1}^{i=47}$$\mid\log_{10}(T_{1/2}(\text{cal}.)(i))-\log_{10}T_{1/2}(\text{exp}.)(i)\mid$/47=1.61,
which means the average deviation between theoretical spontaneous
fission half-live and the experimental ones is less than $10^{2}$
times. This level of agreement is very satisfactory because the
spontaneous fission is much more complex than other decay modes.
\begin{figure}[htbp]
\includegraphics[width=0.50\textwidth]{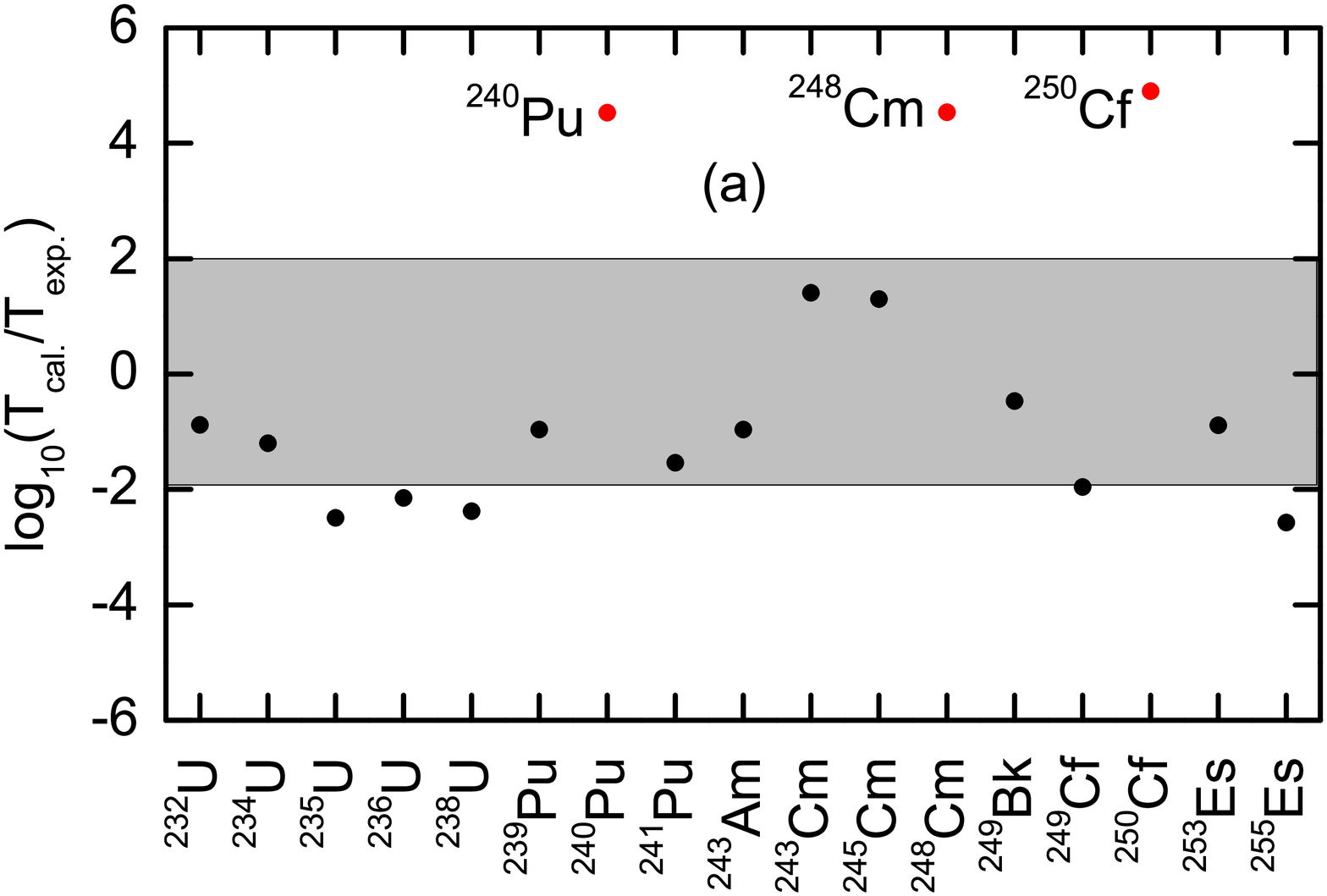}
\includegraphics[width=0.50\textwidth]{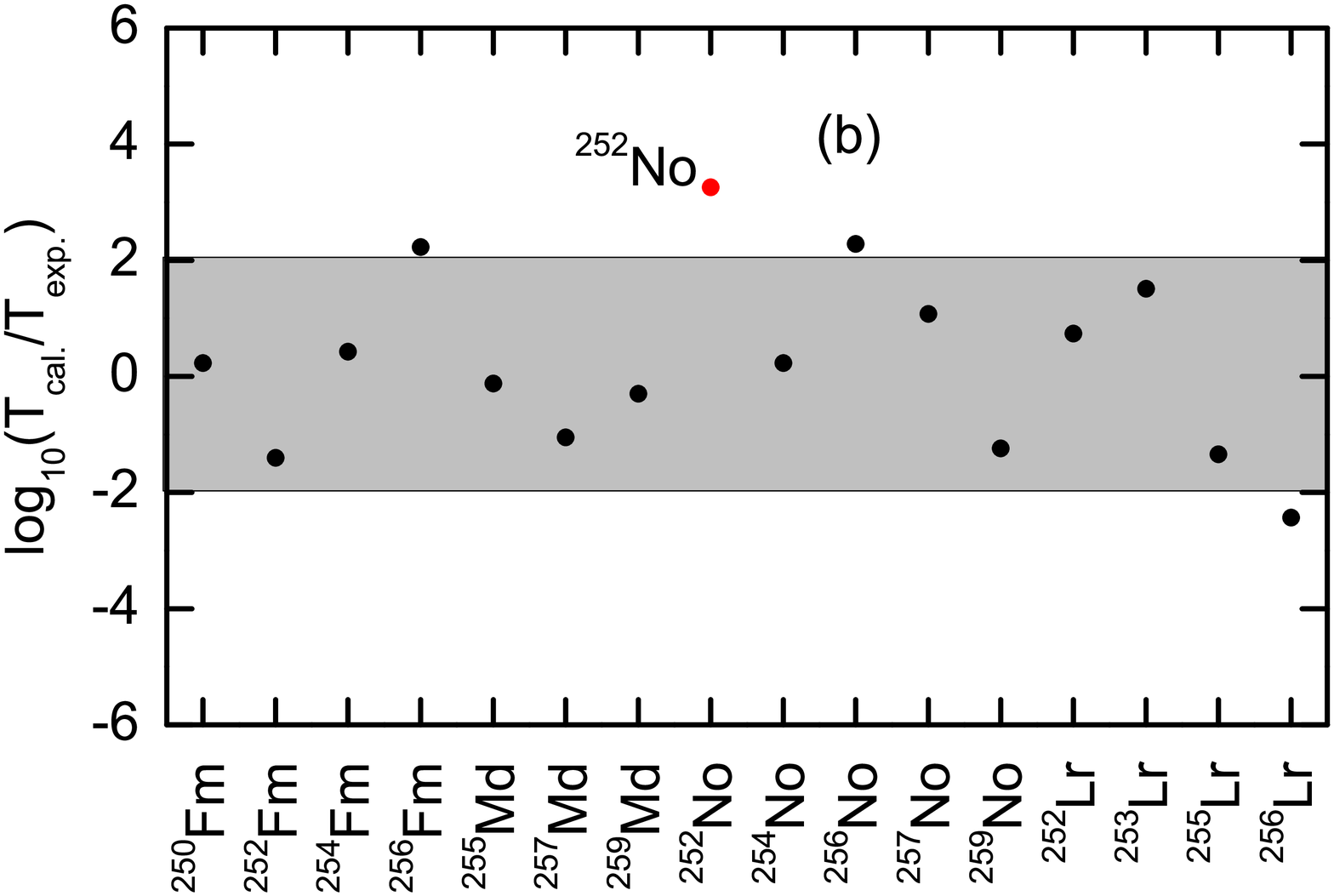}
\includegraphics[width=0.50\textwidth]{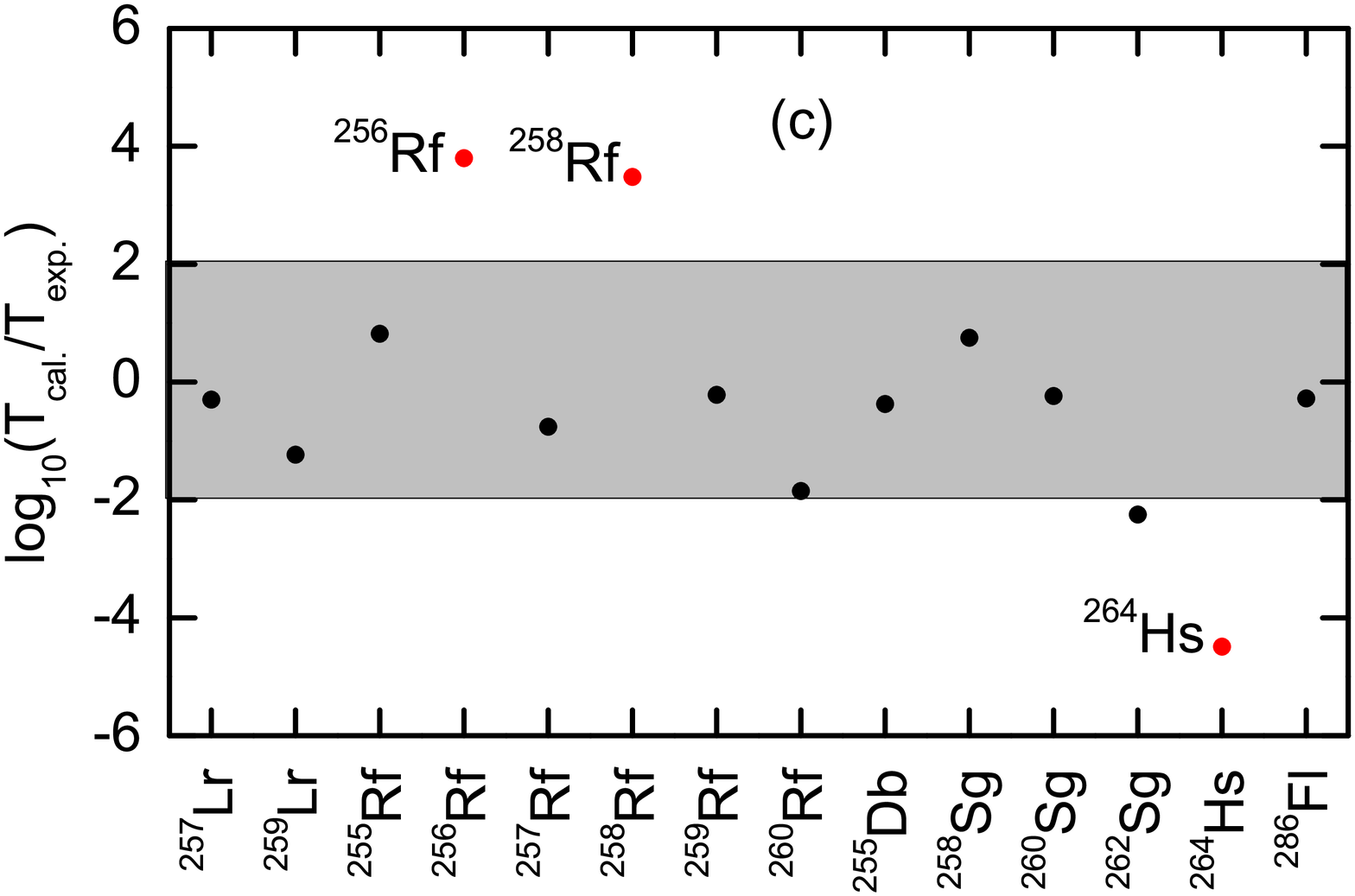}
\caption{\label{Fig2} Deviations between the logarithms of the
calculated half-lives and the experimental values for spontaneous
fission from different parent nuclei.}
\end{figure}

\begin{table}[h]
\label{tab:fis} \caption{Comparison between experimental and
theoretical spontaneous fission half-lives(the unit is seconds) of
heavy and super-heavy nuclei.}
\begin{tabular}{lcclcc}
\hline
Nucleus             &$T_{1/2}$(exp.)   &$T_{1/2}$(the.)       & Nucleus          &$T_{1/2}$(exp.)  & $T_{1/2}$(the.) \\
\hline
$_{92}^{232}$U      &$2.5\times10^{21}$&$3.3\times10^{20}$    &$_{102}^{252}$No  &$1.2\times10^{1}$&$2.2\times10^{4}$\\
$_{92}^{234}$U      &$4.7\times10^{23}$&$2.9\times10^{22}$    &$_{102}^{254}$No  &$3.0\times10^{4}$&$5.1\times10^{4}$\\
$_{92}^{235}$U      &$3.1\times10^{26}$&$1.0\times10^{24}$    &$_{102}^{256}$No  &$1.1\times10^{2}$&$2.1\times10^{4}$\\
$_{92}^{236}$U      &$7.8\times10^{23}$&$5.5\times10^{21}$    &$_{102}^{257}$No  &$1.7\times10^{3}$&$2.0\times10^{4}$ \\
$_{92}^{238}$U      &$2.6\times10^{23}$&$1.1\times10^{21}$    &$_{102}^{259}$No &$3.5\times10^{4}$&$2.0\times10^{3}$\\
$_{94}^{239}$Pu     &$2.5\times10^{23}$&$2.7\times10^{22}$    &$_{103}^{252}$Lr &$3.6\times10^{1}$&$2.0\times10^{2}$\\
$_{94}^{240}$Pu     &$1.5\times10^{18}$&$5.1\times10^{22}$    &$_{103}^{253}$Lr &$2.9\times10^{1}$&$9.2\times10^{2}$\\
$_{94}^{241}$Pu     &$2.3\times10^{24}$&$6.6\times10^{22}$    &$_{103}^{255}$Lr &$2.2\times10^{4}$&$1.0\times10^{3}$\\
$_{95}^{243}$Am     &$6.3\times10^{21}$&$6.9\times10^{20}$    &$_{103}^{256}$Lr &$9.0\times10^{5}$&$3.3\times10^{3}$ \\
$_{96}^{243}$Cm     &$1.7\times10^{19}$&$4.4\times10^{20}$    &$_{103}^{257}$Lr &$2.2\times10^{3}$&$1.1\times10^{3}$\\
$_{96}^{245}$Cm     &$4.4\times10^{19}$&$8.6\times10^{20}$    &$_{103}^{259}$Lr &$5.8\times10^{3}$&$3.4\times10^{2}$\\
$_{96}^{248}$Cm     &$1.3\times10^{14}$&$4.6\times10^{18}$    &$_{104}^{255}$Rf &$3.2\times10^{0}$&$2.1\times10^{1}$\\
$_{97}^{249}$Bk     &$6.1\times10^{16}$&$2.1\times10^{16}$    &$_{104}^{256}$Rf &$6.4\times10^{-3}$&$4.0\times10^{1}$\\
$_{98}^{249}$Cf     &$2.2\times10^{18}$&$2.5\times10^{16}$    &$_{104}^{257}$Rf &$3.9\times10^{2}$&$8.7\times10^{1}$\\
$_{98}^{250}$Cf     &$5.2\times10^{11}$&$4.1\times10^{16}$    &$_{104}^{258}$Rf &$1.4\times10^{-2}$&$4.2\times10^{1}$\\
$_{99}^{253}$Es     &$2.0\times10^{13}$&$2.6\times10^{12}$    &$_{104}^{259}$Rf &$4.0\times10^{1}$&$2.4\times10^{1}$\\
$_{99}^{255}$Es     &$8.4\times10^{10}$&$2.3\times10^{8}$     &$_{104}^{260}$Rf &$5.1\times10^{-2}$&$3.6\times10^{0}$\\
$_{100}^{250}$Fm    &$2.6\times10^{7}$ &$4.5\times10^{7}$     &$_{105}^{255}$Db &$8.0\times10^{-1}$&$3.5\times10^{-1}$\\
$_{100}^{252}$Fm    &$4.0\times10^{9}$ &$1.6\times10^{8}$     &$_{106}^{258}$Sg &$5.2\times10^{-3}$&$2.9\times10^{-2}$\\
$_{100}^{254}$Fm    &$1.9\times10^{7}$ &$5.2\times10^{7}$     &$_{106}^{260}$Sg &$7.2\times10^{-3}$&$4.1\times10^{-3}$\\
$_{100}^{256}$Fm    &$1.0\times10^{4}$ &$1.7\times10^{6}$     &$_{106}^{262}$Sg &$7.0\times10^{-3}$&$3.9\times10^{-5}$\\
$_{101}^{255}$Md    &$1.1\times10^{6}$ &$8.4\times10^{5}$     &$_{108}^{264}$Hs &$1.6\times10^{-3}$&$5.1\times10^{-8}$\\
$_{101}^{257}$Md    &$2.0\times10^{6}$ &$1.8\times10^{5}$     &$_{114}^{286}$Fl &$1.3\times10^{-1}$&$6.9\times10^{-2}$\\
$_{101}^{259}$Md    &$5.8\times10^{3}$ &$2.9\times10^{3}$   \\
\hline
\end{tabular}
\end{table}

To illustrate the agreement between the calculations and the
experimental data clearly, the comparison of the calculated
spontaneous fission half-lives with the experimental data is shown
in Fig.2. The absolute values of
log$_{10}$($T_{1/2}$(cal.)/$T_{1/2}$(exp.)) are generally less
than the factor 2, this means that the experimental spontaneous
fission half-lives are well reproduced. Here the significant
deviations between theoretical calculations and experimental data
occur only for seven nuclei $^{240}$Pu, $^{248}$Cm, $^{250}$Cf,
$^{252}$No, $^{256}$Rf, $^{258}$Rf and $^{264}$Hs. The deviations
for the half-lives of the seven nuclei are 10$^{4}$, 10$^{4}$,
10$^{5}$, 10$^{3}$, 10$^{4}$, 10$^{3}$ and 10$^{5}$. These nuclei
can be separated into three classes. The first class includes
$^{248}$Cm, $^{250}$Cf, $^{252}$No, $^{256}$Rf and $^{258}$Rf, and
the second and three class include, respectively, $^{264}$Hs and
$^{240}$Pu. For $^{248}$Cm, $^{250}$Cf, $^{252}$No, $^{256}$Rf and
$^{258}$Rf the neutron numbers are near N=152. Especially, for
$N=152$ isotones $^{248}$Cm, $^{250}$Cf and $^{256}$Rf, the
deviations are a little larger. Such a large deviation seems due
to the sub-magic neutron shell closure at $N=152$. To illustrate
more clearly that $N=152$ is the sub-magic neutron shell closure,
$\alpha$-decay half-lives have been calculated within a tunnelling
effect through a potential barrier determined by the GLDM and the
WKB approximation. Fig. 3 represents the plot connecting
calculated $\alpha$-decay half-lives against neutron number of the
even-even parent isotopes with Z ranging from 98 to 108. Two peaks
appear at $N=152$ and $N=162$. In $\alpha$ and cluster emission it
is found that half-life has the minimum value for those decays
which lead to doubly magic daughter
\cite{R38W94,Poe12,Qi09,Sob11}. Therefore the peaks at $N=152$ and
$N=162$ indicate the presence of shell closures at these values.
We would like to point out that many authors
\cite{P39W94,R40S95,G41A96,L42S01,D43N06} have predicted sub-magic
neutron shell closures at $N=152$ and $N=162$. The deviation for
$^{264}$Hs seems indicate that there is a proton shell closure for
$Z=108$. The macroscopic-microscopic model has predicted that
$^{270}$Hs is a deformed doubly magic nucleus \cite{Z43A89}.
Recently, shells at $N=162$ neutrons and $Z=108$ protons were
predicted by GLDM \cite{H44F12}. These predictions have been
supported by experiments \cite{Y44A94,Y45A95,J46D06}. The
spontaneous fission process is complex and there are large
uncertainties existing in the fission process, so one may consider
that the slightly larger deviations of $10^{4}$ for few nuclei are
acceptable\cite{San09}.

\begin{figure}[htbp]
\begin{center}
\includegraphics[width=0.90\textwidth]{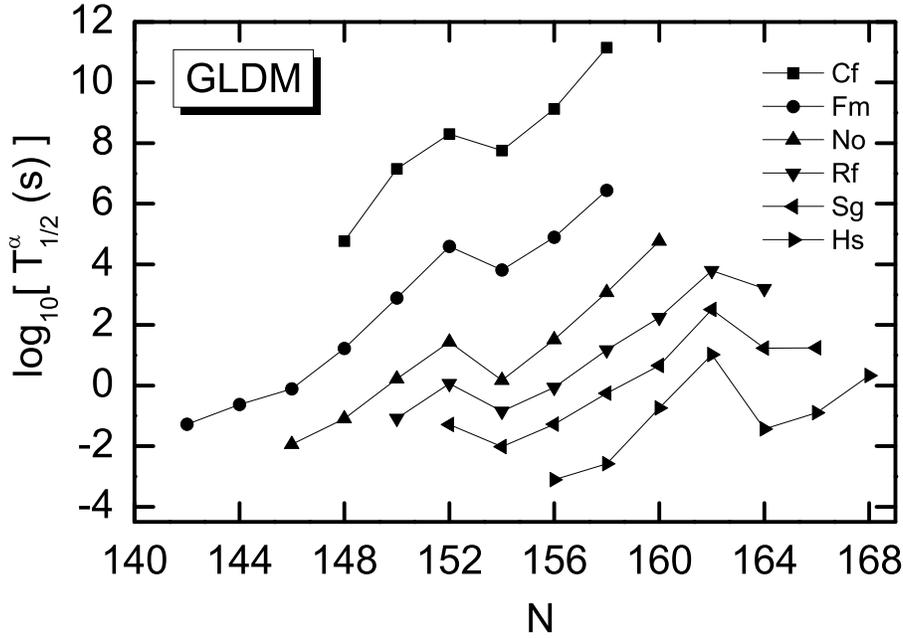}
\caption{\label{Fig3}Theoretical half-lives versus neutron number
for various parent nuclei with $Z=98-108$ emitting
$\alpha$-decay.}
\end{center}
\end{figure}

Fig.4 represents the comparison of our spontaneous fission
half-lives from $^{250}$Fm to $^{262}$Sg with the results from
Ref.\cite{P11J89} and with the experimental values \cite{J44K05}.
Solid triangles represent experimental data and solid circles
denote the present calculated half-lives. Open circles show the
results from Ref. \cite{P11J89}, in which the
Yukawa-plus-exponential model for the macroscopic part of the
potential energy and the Strutinsky shell correction, based on the
Woods-Saxon single-particle potential, for the microscopic part
are employed. The trend of theoretical results from
Ref.\cite{P11J89} follows well the experimental ones. However,
these values \cite{P11J89} are systematical larger than the
experimental ones and larger than ours by up to about six orders
of magnitude. The difference seems to originate mainly from the
shell corrections. In our approach the shell corrections have been
introduced as defined in the Droplet Model \cite{W31D77} with an
attenuation factor. Using this approach, shell corrections only
play a role near the ground state of the compound nucleus and not
at the saddle-point. It has been clearly demonstrated within a
single-particle model with pairing corrections
\cite{W35N70,W36N72} that, for two separated spheroids, the shell
effects are strongly diminished since they are properties of
valence nucleons and that the orbital of which are strongly
perturbed by the nuclear proximity potential. Thus, as soon as the
shape is creviced, the application of the standard shell
corrections to the liquid drop model energy seems to overestimate
the veritable shell effects which are partially destroyed by the
proximity forces. By comparing the present results of spontaneous
fission half-lives with the results from Ref. \cite{P11J89}, we
would like to point out that our calculated values better
reproduce the experimental data.

\begin{figure}[htbp]
\includegraphics[width=0.90\textwidth]{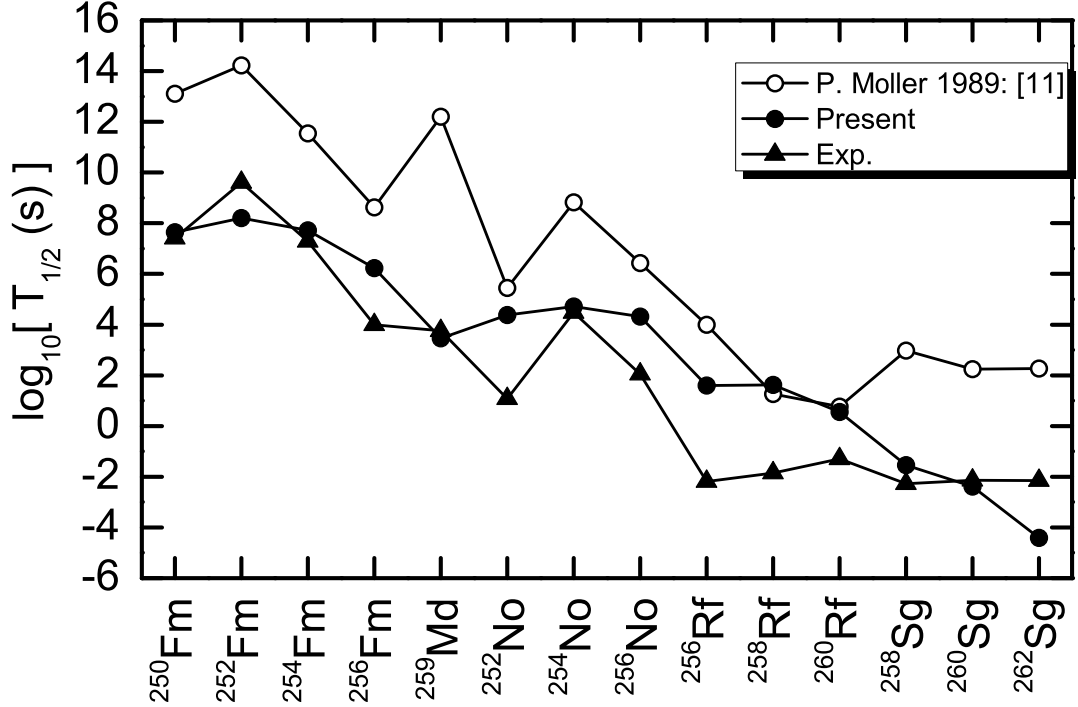}
\caption{\label{Fig4}The comparison of the logarithm of the
present spontaneous fission half-lives with the values taken from
P.M\"{o}ller et al \cite{P11J89} and the experimental data. }
\end{figure}

Fig.5 represents the comparison of our spontaneous fission
half-lives for partial odd-A and odd-odd nuclei with the results
from Ref. \cite{Z37C05} and with the experimental values
\cite{J44K05}. Solid triangles represent the calculated
spontaneous fission half-lives from Ref. \cite{Z37C05}, in which
the phenomenological formula for spontaneous fission half-lives is
given by
\begin{eqnarray}
\log_{10}(T_{1/2}/yr)=21.08+C_{1}\frac{(Z-90-\upsilon)}{A} \\
\nonumber
+C_{2}\frac{(Z-90-\upsilon)^{2}}{A}+C_{3}\frac{(Z-90-\upsilon)^{3}}{A}
\\ \nonumber +C_{4}\frac{(Z-90-\upsilon)(N-Z-52)^{2}}{A},
\end{eqnarray}
where C$_{1}$=-548.825021, C$_{2}$=-5.359139, C$_{3}$=0.767379 and
C$_{4}$=-4.282220, the seniority term $\upsilon$ is $\upsilon$=0
for the spontaneous fission of even-even nuclei and $\upsilon$=2
for spontaneous fission of odd A and odd-odd nuclei. The value
$\upsilon$ indicates the blocking effect of unpaired nucleon on
the transfer of many nucleon-pairs during the fission process. The
agreement between the experimental
 and theoretical half-lives are generally good except for a
few cases (e.g. $^{259}$Rf and $^{255}$Db).

\begin{figure}[htbp]
\includegraphics[width=0.90\textwidth]{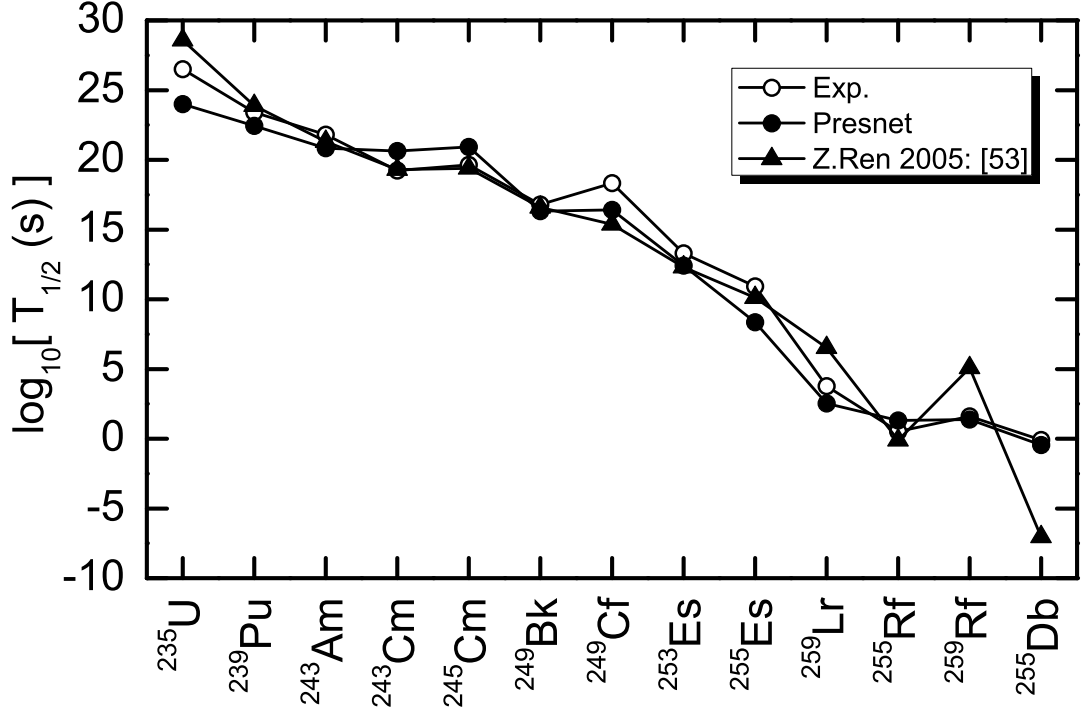}
\caption{\label{Fig5}The comparison of the logarithm of the
present spontaneous fission half-lives with the values taken from
Ren et al \cite{Z37C05} and the experimental data.}
\end{figure}

Since the present calculated half-lives agree well with the
experimental ones, the calculations are extended to provide some
predictions for spontaneous fission half-lives, which will be
useful for future experiments to synthesize and detect the new
SHN. The predictions are shown for Z=114-120 isotopic chains in
Table \ref{tab:pre}. For some SHN, the spontaneous fission
half-lives are long enough to be measured with the present
experimental setups.

\section{Summary}\noindent

The spontaneous fission process in the quasimolecular shape valley
is investigated within a generalized liquid drop model where the
microscopic shell corrections and the phenomenological pairing
corrections are considered. A systematic calculation on
spontaneous fission half-lives for heavy and superheavy nuclei
with proton number Z~$\geq$~92 is performed. The calculated
half-lives are in good agreement with the experimental data. For
most nuclei, the experimental half-lives are reproduced within a
factor of $10^{2}$. This level of agreement is very satisfactory
because the spontaneous fission is much more complex than other
decay modes such as cluster radioactivities and $\alpha$-decay.
Spontaneous fission half-lives of the isotopes of $Z=114-120$ are
predicted, presuming that this might help to discriminate between
all the possible future experiments.

\begin{table}[h]
\label{tab:pre} \caption{Predicted half-lives (the unit is
seconds) of spontaneous fission of even-even nuclei on Z=114-120
isotopic chain.}
\begin{tabular}{clclclcl}
\hline
Nucleus & $T_{1/2}$(cal.) & Nucleus &
 $T_{1/2}$(cal.) &Nucleus &
$T_{1/2}$(cal.) & Nucleus & $T_{1/2}$(cal.)\\
\hline
$^{280}Fl$&$2.2\times10^{-3}$    &$^{282}Fl$&$2.0\times10^{-1}$  &$^{284}Fl$&$1.2\times10^{-1}$&
$^{288}Fl$&$1.1\times10^{0}$  \\  $^{290}Fl$&$2.3\times10^{3}$    &$^{292}Fl$&$6.9\times10^{7}$    &$^{294}Fl$&$3.6\times10^{8}$&
$^{296}Fl$&$3.3\times10^{8}$  \\   $^{300}Fl$&$2.6\times10^{-1}$  &$^{286}Lv$&$1.4\times10^{-5}$   &$^{288}Lv$&$2.0\times10^{-5}$&
$^{290}Lv$&$2.5\times10^{-3}$ \\    $^{292}Lv$&$5.5\times10^{0}$  &$^{294}Lv$&$1.1\times10^{5}$    &$^{296}Lv$&$2.7\times10^{5}$&
$^{298}Lv$&$8.8\times10^{3}$  \\   $^{300}Lv$&$1.1\times10^{4}$  &$^{302}Lv$&$2.6\times10^{-5}$    &$^{292}118$&$1.4\times10^{-4}$  &
$^{294}118$&$2.2\times10^{-1}$ \\   $^{296}118$&$3.9\times10^{3}$ &$^{298}118$&$1.4\times10^{2}$      &$^{300}118$&$2.8\times10^{1}$&
$^{302}118$&$6.9\times10^{1}$ \\   $^{294}120$&$6.3\times10^{-5}$  &$^{296}116$&$8.6\times10^{-2}$     &$^{298}120$&$1.1\times10^{1}$ &
$^{300}120$&$1.0\times10^{1}$  \\   $^{302}120$&$2.3\times10^{1}$  &$^{304}120$&$3.1\times10^{1}$     &$^{306}120$&$3.0\times10^{-6}$&\\
\hline
\end{tabular}
\end{table}

\section*{Acknowledgements}
The work is supported by the Natural Science Foundation of China
(Grants 10775061, 11120101005, 11105035, 10975064 and 11175074),
the Fundamental Research Funds for the Central Universities
(grants lzujbky-2012-5), by the CAS Knowledge Innovation Project
NO.KJCX-SYW-N02.

\renewcommand{\baselinestretch}{1.0}

\bigskip
\end{document}